\documentclass{article}
\begin{document}
\title{Non-linear connections on phase space and the Lorentz force law}
\author{John H. van Drie\\
        Meridien Research\\ 
        Kalamazoo, MI  USA\\
        www.johnvandrie.com}
\maketitle
\begin{abstract}
The equations of parallel transport for a non-linear connection on phase
space are examined.  It is shown that, for a free-particle Lagrangian, 
the connection term first-order in momentum reproduces the geodesic
equation of General Relativity and the term zeroth-order
in the momentum reproduces the Lorentz electromagnetic
force law.  Hence from one mathematical expression, a non-linear parallel 
transport equation, one can derive the interaction laws for both the
gravitational and electromagnetic forces.  These equations are free of the 
difficulties associated with formalism of Weyl, which forms the basis for
the theory of Yang and Mills. 
\end{abstract}
\section{Introduction}
After Einstein's description of the gravitational force
as a manifestations of the curvature of space, numerous methods were
proposed for a similar, geometric explanation of the Lorentz force
law [1-4].  Weyl's proposal [1] is especially significant, as it is the
basis for the field equations of Yang-Mills [5], the basis of modern field
theory.  Wu and Yang [6] have noted, however, that the Weyl/Yang-Mills
approach leads to certain
mathematical difficulties, when viewed in its original differential-geometric
context.  
Presented here is another differential-geometric method for deriving the 
Lorentz electromagnetic force law 
which is free of these difficulties.  The method described here
develops a differential-geometric formalism using relativistic phase 
space as the underlying
structure (unlike General Relativity and Weyl's theory, which use 
spacetime and its tangent space as the underlying structure).  This method
provides a conceptually clearer understanding of the relationship between
the physical quantity of Newtonian {\bf force}, and the mathematical quantity
of {\bf connection}.  The connection under consideration is a more general type
of connection than that of Riemannian geometry, the mathematical basis of
General Relativity.  Rather than considering {\it geodesics}, i.e. paths
of minimum distance (the distance determined by the metric), the method here
uses the equations of {\it parallel transport}, which under appropriate 
constraints reduce to the geodesic equation.  
The relaxation of these constraints gives rise to new mathematical terms, 
one of which it is shown may be interpreted physically as the electromagnetic 
field.
\section{The geometric structure of phase space}
Phase space is the natural setting for studying the dynamics of an 
N-particle system.  The purpose of this section is to describe the
{\it geometric} structure of phase space, and establish in a general
way the relation between the mathematical, geometric quantity ``connection''
with the physical quantity ``force''.
Given 4-dimensional spacetime and a particle whose position at a point in 
proper time $\tau$ may be characterized by the coordinates 
$\{ x^{\mu}(\tau) \}_{\mu=0}^3$,
and a Lagrangian $L(x^{\mu},\frac{dx^{\mu}}{d\tau},\tau)$ describing this
physical system, one defines momenta as
\begin{eqnarray*}
             p_i & = & \frac{\partial L}{\partial \dot{x}^i}\\
             p_0 & = & \sum_i p_i \dot{x}^i - L
\end{eqnarray*}
where i ranges over the 3 spatial coordinates, 0 denotes the time
coordinate, and $\dot{x}^i = \frac{dx^i}{dx^0}$.  
One constructs the phase space $\Pi$ , such that each point
in $\Pi$ is described by coordinates $(x^{\mu},p_{\mu})$.  We will consider
two infinitesimally displaced points in spacetime, ${\bf x}$ and 
${\bf x'}$, and will define $\Pi_x$ as the subspace of $\Pi$ representing
all possible momenta the particle may take at a point $\bf{x}$.  If,
at a time $\tau$, the particle is at point $\bf{x}$, we may associate with it
some momentum, i.e. some element $p \in \Pi_x$.  Similarly, if at a time
$\tau + d\tau$, the particle is at $\bf{x'}$, we may associate it with some
$p' \in \Pi_{x'}$.  In the classical Newtonian view, the change in momentum
over this interval of time $d\tau$ represents the forces acting on the
particle.
However, viewing this same situation geometrically, we see that
a subtle point has been overlooked in this classical Newtonian view:
\begin{quote}
      there is no way {\it a priori} to associate some element 
      $p \in \Pi_x$ with some $p' \in \Pi_{x'}$, since {\bf x} and
      {\bf x'} are infinitesimally displaced.
\end{quote}
{\it We must first parallel transport} $p'$ {\it back to} $\Pi_x$, {\it and
then compute the difference between it and} $p \in \Pi_x$.  Thus,
{\it we must define a} {\bf connection} {\it before we can construct the
rate of change of momentum representing the Newtonian force}.
How does one choose a connection?  Newtonian mechanics is usually based on
a Euclidean connection, with forces described by terms in the Lagrangian.
Instead, we will consider only free-particle Lagrangians, and 
construct connections such that the parallel-transported change in momentum
is zero along a particle's path.  In other words, rather than ascribing
differences between $p$ and $p'$ to an external force (i.e. forces cause
deviations from Euclidean straightness), we will ascribe the difference
between $p$ and $p'$ to the parallel transport of $p \in \Pi_x$ to
$p' \in \Pi_{x'}$.  Loosely speaking, {\it we shall describe ``forces''
in terms of connections}.
Conceptually, this is identical to Einstein's programme in the
General Theory of Relativity.  This identification of force with connection
allows one to identify stress-energy with curvature, where mathematical
identities on the curvature automatically produce the necessary conservation
theorems on stress-energy [7].  However, the approach described here generalizes
Einstein's 
approach, by going beyond the affine connections derived from a 
metric in Riemannian geometry, to the most general expression for a non-linear
connection on phase space.  Whereas in Riemannian geometry, the metric is the
fundamental quantity, from which quantities such as the connection and 
curvature are derived, these connections we consider are the most 
fundamental quantity from which other
geometric quantities such as curvature are derived [8].  Such connections
may not in general be derived from a metric.  By adopting
this more general mathematical framework, we can ``geometrize'' a wider
variety of forces, notably the electromagnetic force.
\section{Linear connections on phase space}
An arbitrary connection on phase space may be defined by a Pfaffian 
ideal of 1-forms $\theta_{\mu}$ [9],
\begin{equation}
      \theta_{\mu} = dp_{\mu} - f_{\mu \nu}(x,p)dx^{\nu}  \label{eq:parxport}
\end{equation}
where the $f_{\mu \nu}$ are arbitrary functions of position ${\bf x}$ 
and momentum p.  (We use here a pseudo-Euclidean metric $g_{\mu \nu} =
\eta_{\mu \nu}, \eta_{0 0}=-1, \eta_{1 1} = \eta_{2 2} = \eta_{3 3} = 1$,
0 otherwise).  We shall first consider the homogeneous-linear connection
$f_{\mu \nu}(x,p)=h_{\mu \nu \alpha}(x) p^{\alpha}$, where 
$h_{\mu \nu \alpha}$ are arbitrary functions of position.  Defining the
connection 1-forms as
$$
           \omega_{\mu \alpha} = h_{\mu \nu \alpha}(x) dx^{\nu}
$$
(equivalent to Cartan's connection 1-forms), equation~\ref{eq:parxport}
becomes
\begin{equation}
         \theta_{\mu} = dp_{\mu} - \omega_{\mu \alpha} p^{\alpha} \label{eq:hlxport}
\end{equation}
for this homogeneous-linear connection.  We can now calculate explicitly the
equations for the curves defined by this Pfaffian ideal.  Denoting by
{\bf u} the unit tangent vector to these curves, we know that
${\bf u} \rfloor \theta_{\mu} = 0$ for all $\mu$, where $\rfloor$ 
denotes contraction of a vector with a 1-form.  Equation~\ref{eq:hlxport}
then becomes
$$
     {\bf u} \rfloor \theta_{\mu} = 0  =  {\bf u} \rfloor dp_{\mu}
                  - {\bf u} \rfloor (\omega_{\mu \alpha} p^{\alpha})
$$
.
$$           
      0  =  u^{\nu}\frac{\partial p_{\mu}}{\partial x^{\nu}} -
                 u^{\nu}(h_{\mu \nu \alpha} p^{\alpha})
$$
.
\begin{equation}                 
      0  =  \frac{dp_{\mu}}{d\tau} - 
           h_{\mu \nu \alpha}p^{\alpha} u^{\nu} \label{eq:hxport2}
\end{equation}
where $\tau$ is a parametrization of the curve, and where 
$\frac{dp_{\mu}}{d\tau}$
denotes the derivative along the curve, and where we have taken
$$
    \frac{dp_{\mu}}{d\tau} = {\bf u} \rfloor dp_{\mu} = 
              u^{\alpha}\frac{\partial p_{\mu}}{\partial x^{\alpha}}               
$$              
Assuming a free point-particle Lagrangian, with mass m (a scalar constant),
for which [10],
$$
        p_{\mu} = m u_{\mu}
$$
equation~\ref{eq:hxport2} becomes
$$
     0 = m \frac{du_{\mu}}{d\tau} - mh_{\mu \nu \alpha}u^{\alpha}u^{\nu},
$$
or,
\begin{equation}
       \frac{du_{\mu}}{d\tau} - h_{\mu \nu \alpha} u^{\alpha} u^{\nu} = 0.
           \label{eq:hlxport3}
\end{equation}
If one associates the coefficient $h_{\mu \nu \alpha}$ for our homogeneous-
linear connection with $\Gamma_{\mu \nu \alpha}$, the affine connection of
Riemannian geometry,
$$
    h_{\mu \nu \alpha} = - \Gamma_{\mu \nu \alpha}
$$
equation~\ref{eq:hlxport3} becomes the ``geodesic equation'' for a
Riemannian space [11],
$$
          \frac{du_\mu}{d\tau} + \Gamma_{\mu \nu \alpha}u^{\alpha}u^{\nu} = 0.
$$                      
Einstein demonstrated that this equation describes the forces of gravitation,
once the curvature of space is related to stress energy, $T_{\mu \nu}$[12]:
$$
     T_{\mu \nu} = R_{\mu \nu} - \frac{1}{2} g_{\mu \nu} R
$$
where
$$
     R = g^{\mu \nu} R_{\mu \nu},
$$
and
$$
   R_{\mu \nu} = \Gamma^{\alpha}_{\mu \nu , \alpha}
               - \Gamma^{\alpha}_{\mu \alpha , \nu}
               + \Gamma^{\alpha}_{\beta \alpha} \Gamma^{\beta}_{\mu \nu}
               - \Gamma^{\alpha}_{\beta \nu} \Gamma^{\beta}_{\mu \alpha}.
$$
where $,\alpha$ denotes derivation, i.e.
$$
  \Gamma^{\alpha}_{\mu \alpha ,\nu} = 
          \frac{\partial \Gamma^{\alpha}_{\mu \alpha}}{\partial x^{\nu}}.
$$          
\section{Non-linear connections on phase space}
\subsection{Derivation of parallel transport equation on phase space}
We now return to equation~\ref{eq:parxport}, solutions to which describe
the paths of parallel transport in phase space, to relax the assumption
of homogeneous-linearity which led to equation~\ref{eq:hlxport}.  Now
we shall consider connections of the form
$$
f_{\mu \nu}(p,x) = \stackrel{(0)}{h}_{\mu \nu}(x)
                 + \stackrel{(1)}{h}_{\mu \nu \alpha}(x) p^{\alpha}
$$
where the $\stackrel{(k)}{h}_{\mu \nu \alpha_1 \dots \alpha_k}(x)$ are
functions only of x.  Defining the set of connection 1-forms
$$
   \stackrel{(0)}{\omega}_{\mu}  = \stackrel{(0)}{h}_{\mu \nu} dx^{\nu}
$$
.
$$   
   \stackrel{(1)}{\omega}_{\mu \alpha}  =  
              \stackrel{(1)}{h}_{\mu \nu \alpha} dx^{\nu}
$$
equation~\ref{eq:parxport} becomes
$$
 \theta_{\mu} = dp_{\mu}  
             - \stackrel{(0)}{\omega}_{\mu}
             - \stackrel{(1)}{\omega}_{\mu \alpha} p^{\alpha}
$$
As before, we explicitly calculate the equations for the curves which annul 
this Pfaffian ideal, by contracting the unit tangent vector {\bf u} with
all of the 1-forms $\theta_{\mu}$:
$$
    {\bf u} \rfloor \theta_{\mu} =  0
$$
.
$$
 0    =    {\bf u} \rfloor dp_{\mu} 
       -  {\bf u} \rfloor \stackrel{(0)}{\omega}_{\mu}
       -  {\bf u} \rfloor \stackrel{(1)}{\omega}_{\mu \alpha} p^{\alpha}
$$
Recalling our earlier use of the derivative along the curve, and
explicitly writing the contraction, this becomes
$$
  0 = \frac{dp_{\mu}}{d\tau} 
            - u^{\gamma} \stackrel{(0)}{h}_{\mu \gamma}
            - u^{\gamma} \stackrel{(1)}{h}_{\mu \gamma \alpha} p^{\alpha}
$$
Assuming again a point-particle Lagrangian, $p_{\mu} = m u_{\mu}$, this
becomes
$$
  0 = m\frac{du_{\mu}}{d\tau} 
               - u^{\gamma} \stackrel{(0)}{h}_{\mu \gamma}
             - m u^{\gamma} \stackrel{(1)}{h}_{\mu \gamma \alpha} u^{\alpha}
$$
or, in a form where the mass-dependence of the different terms is explicit,
\begin{equation}
  0 = \frac{du_{\mu}}{d\tau} 
            - \frac{1}{m} \stackrel{(0)}{h}_{\mu \gamma} u^{\gamma}
            - \stackrel{(1)}{h}_{\mu \gamma \alpha} u^{\gamma} u^{\alpha} \label{eq:master}
\end{equation}
Notice that, expressed in this form, {\it the equivalence principle
is manifest}.  Only one term may be without a mass term, the first-order term.
Mathematically, this must describe paths which particles will follow
independent of their mass, which physically corresponds to gravitation,
as Einstein's General theory shows.  Other geometrically-derived
forces, the zeroth-order term, and 2nd-order terms and higher, correspond
to forces whose effects will be dependent on the mass of the particle.
\subsection{Physical interpretation of the lowest order term of 
this parallel transport equation}
In order to interpret equation~\ref{eq:master}, we will associate the 
connection term $\stackrel{(0)}{h}_{\mu \gamma}$, which is independent of
momentum but a function of position, with the Faraday tensor of
electromagnetism [13],
$$
F_{\mu \gamma} =
\left( \begin{array}{clcr}
     0       & E_x        & E_y        &E_z    \\
   -E_x      & 0          & B_z        &-B_x   \\
   -E_y      & -B_z       & 0          & B_y   \\
   -E_z      & B_x        & -B_y       & 0  
\end{array} \right)     \label{eq:faraday}
$$
where {\bf E} = $(E_x,E_y,E_z)$ is the electric-field spatial 3-vector,
and {\bf B} = $(B_x,B_y,B_z)$ is the magnetic-field spatial 3-vector.
Combining 
$$
    \stackrel{(0)}{h}_{\mu \gamma} = eF_{\mu \gamma}
$$
where e is the electric charge, with the definition of ${\bf F}$  and 
equation~\ref{eq:master}, and setting the first and second order terms,
$\stackrel{(1)}{h_{\mu \gamma \alpha}}$ and 
$\stackrel{(2)}{h_{\mu \gamma \alpha \beta}}$ to zero, we are led
to
\begin{equation}
   \frac{d}{dt}(m{\bf v}) = e(\bf{E} - \bf{v} \times {\bf B}) \label{eq:lorentz}
\end{equation}
where {\bf v} is the velocity spatial 3-vector, and $\frac{d}{dt}$
denotes the derivative with respect to coordinate time.  This equation
expresses the time-rate-of-change of the Newtonian momentum, $m {\bf v}$,
to a term which has precisely the form of the Lorentz force law.  Hence
by starting with equation~\ref{eq:master}, which was derived mathematically
from the expression for a non-linear connection on phase space 
(equation~\ref{eq:parxport}), we are led to a physical equation, the Lorentz force
law, whose validity is derived empirically.  Therefore, this development 
may be interpreted as a geometric derivation of the Lorentz force law. 
If we repeat this derivation, one sees
that equation~\ref{eq:master} describes the path of a particle in a 
Riemannian connection 
$\Gamma_{\mu \nu \alpha} = - \stackrel{(1)}{h}_{\mu \gamma \alpha}$ in
the presence of an electromagnetic field [14].
$$
 \frac{du^{\alpha}}{d\tau} + \Gamma^{\alpha}_{\mu \nu} u^{\mu} u^{\nu}
      = \frac{e}{m} F^{\alpha}_{\beta} u^{\beta}
$$      
\subsection{Other aspects of the physical interpretation of the
non-linear parallel transport law}
We have shown how the classical electrodynamic interaction law
may be derived from the mathematics of non-linear connections on phase
space.  A proper understanding of the additional terms present
in this expression would require a quantum-mechanical formulation
of this classical law.  Recall that the quantum-mechanically, the
electromagnetic interaction is introduced by applying
the ``minimal substitution'' $ {\bf p} \rightarrow {\bf p} - e{\bf A}$
to the path equations in the absence of an electromagnetic field[15], 
where {\bf A}
is the vector potential, whose exterior derivative is the Faraday tensor
${\bf F = dA} [16]$.  However, in our derivation via the non-linear parallel
transport law, we have lost the simple understanding of the vector potential.
Hence, a straightforward application of equation~\ref{eq:master} 
to quantum mechanics cannot
be made.  Possibilities for making the application to quantum mechanics,
by looking for the equations of parallel transport of $|\xi \rangle$ which
provide the correct equations of parallel transport for  
$p_{\mu} = 
- i \hbar \langle \xi | \frac{\partial}{\partial x^{\mu}} | \xi \rangle$,
where $\xi$ is the 4-component spinor of the Dirac equation, will be
examined separately.
One important qualitative feature of a quantum-mechanical theory of the 
electromagnetic interaction based on this approach is already apparent:
we make the following associations between mathematical and physical
quantities
$$
\begin{array}{clcr}
 {\bf connection}    &&  {\bf force}  \\
  \downarrow d       &&  \downarrow d \\
 {\bf curvature}     &&  {\bf sources}\\
  \downarrow d       &&  \downarrow d \\
       0             &&      0        \\
 (Bianchi \; identity)  && (conservation \; of \; energy) 
\end{array}
$$
exactly as can be made in Einstein's General theory[7].  This is in contrast
to the geometric derivation of the electromagnetic interaction law
devised by Weyl [1],  as has been noted by Wu and Yang [6] (this table is 
taken from their Table I):
$$
\begin{array}{clcr}
      {\bf connection}   && {\bf gauge \; potential}  \\
       \downarrow d      && \downarrow d           \\
      {\bf curvature}    &&  {\bf field \; strength}  \\
       \downarrow        && \downarrow d           \\
            ?            &&  {\bf sources} 
\end{array}
$$
\section{Acknowledgements}
This work was begun while the author was a National Science Foundation
Fellow at the California Institute of Technology, and was continued as
a Belgian-American Educational Foundation Fellow at the 
Universit\'{e} Libre de Bruxelles.
\newpage
\section{References}
\begin{flushleft}
[1] H. Weyl, {\it Z. Phys.}, {\bf 56}, 330 (1929).

[2] T. Kaluza, {\it Sitzungsb. preuss. akad. Wiss.}, {\bf 966} (1921).

[3] O. Klein, {\it Z. Phys.}, {\bf 37}, 895 (1926).

[4] V. Hlavaty, {\it Geometry of Einstein's Unified Field Theory}, Groningen,
1957.

[5] C. N. Yang and R. L. Mills, {\it Phys. Rev.}, {\bf 96}, 191 (1954).

[6]  T. T. Wu and C. N. Yang, {\it Phys. Rev. D}, {\bf 12}(12), 3845 (1975).

[7]  C. Misner, K.Thorne, and J. Wheeler, {\it Gravitation}, San Francisco:
W. H. Freeman, 1973, Chapter 15.

[8]  W. Slebodzinski, {\it Exterior Forms and their Applications}, Warsaw:
Polish Scientific Publishers, 1970, chapter IX.

[9]  R. Hermann, {\it Gauge Fields and Cartan-Ehresmann Connections}, Brookline,
MA:  Math-Sci Press, 1975, p. 110, equation 3.5.  This work details the
significance of the multiple connection forms which appear in the text.

[10]  Ref. 7, p. 201.

[11]  {\it ibid.}, p. 224.

[12]  {\it ibid.}

[13]  {\it ibid.}, p. 73.

[14]  {\it ibid.}, eq. 20.41.

[15]  R. P. Feynman, {\it Quantum Electrodynamics}, Reading, MA:  W. A. 
Benjamin, 1962, p. 4.

[16]  Ref. 7, p. 569.  Note that the minimal substitution
works classically as well:  
another way to ``derive'' equation (20.41) in [7] is to take the usual 
geodesic equation, 
equation~\ref{eq:hlxport3}, and apply the ``minimal substitution'',
${\bf p} \rightarrow {\bf p} -e{\bf A}$, or equivalently ${\bf dp} \rightarrow
{\bf dp} - e{\bf F}$.
\end{flushleft}
\end{document}